# Effect of strain relaxation on magnetotransport properties of epitaxial $La_{0.7}Ca_{0.3}MnO_3$ films

**P. K. Siwach[a] , H. K. Singh[b] and O. N. Srivastava[a #]**

**a** Physics Department, Banaras Hindu University, Varanasi-221005, India.
**b** National Physical Laboratory, Dr K S Krishnan Road, New Delhi-110012, India.

## Abstract

In this paper, we have studied the effect of strain relaxation on magneto-transport properties of $La_{0.7}Ca_{0.3}MnO_3$ epitaxial films (200 nm thick), which were deposited by pulsed laser deposition technique under identical conditions. All the films are epitaxial and have cubic unit cell. The amount of strain relaxation has been varied by taking three different single crystal substrates of $SrTiO_3$, $LaAlO_3$ and MgO. It has been found that for thicker films the strain gets relaxed and produces variable amount of disorder depending on the strength of strain relaxation. The magnitude of lattice relaxation has been found to be 0.384, 3.057 and 6.411 percent for film deposited on $SrTiO_3$, $LaAlO_3$ and MgO respectively. The films on $LaAlO_3$ and $SrTiO_3$ show higher $T_{IM}$ of 243 K and 217 K respectively as compared to $T_{IM}$ of 191 K for the film on MgO. Similarly $T_C$ of the films on $SrTiO_3$ and $LaAlO_3$ is sharper and has value of 245 K and 220 K respectively whereas the $T_C$ of the film on MgO is 175 K. Higher degree of relaxation creates more defects and hence $T_{IM}$ ($T_C$) of the film on MgO is significantly lower than of $SrTiO_3$ and $LaAlO_3$. We have adopted a different approach to correlate the effect of strain relaxation on magneto-transport properties of LCMO films by evaluating the resistivity variation through Mott's VRH model. The variable presence of disorder in these thick films due to lattice relaxation which have been analyzed through Mott's VRH model provides a strong additional evidence that the strength of lattice relaxation produces disorder dominantly by increase in density of defects such as stacking faults, dislocations, etc. which affect the magneto-transport properties of thick epitaxial $La_{0.7}Ca_{0.3}MnO_3$ films.

[#] Corresponding Author
Electronic mail: hepons@yahoo.com

## 1. Introduction

The discovery of the colossal magnetoresistance (CMR) effect in epitaxial manganite thin films has renewed interest in these materials for technological usefulness such as potential sensor and magnetic recording applications as well as the need to understand the mechanisms underlying their behavior[1–3]. In epitaxial films, biaxial strain has been reported to have very strong effect on the changes in $T_C$ and insulator-metal transition temperature ($T_{IM}$)[4-23]. Several publications have reported that both $T_C$ and $T_{IM}$ can vary with film thickness[14–23]. In most cases, these changes are interpreted in terms of substrate-induced strain, which relaxes with increase in thickness[18–23]. The lattice mismatch δ along the interface is defined by $\delta = (a_{p\ substrate} - a_{p\ bulk})/a_{p\ substrate}$. When the film is grown on a substrate whose lattice parameter is smaller or larger than that of the bulk material, the epitaxial strain is expected to be compressive (the cell is elongated along the growth direction and compressed in the film's plane) or tensile (the cell is elongated in the film's plane and compressed along the out-plane growth direction), respectively. Compressive strain usually reduces the resistivity and shifts $T_C$ towards higher temperature. These effects have been confirmed in $La_{0.7}Ca_{0.3}MnO_3$ films [6,7] and $La_{0.7}Sr_{0.3}MnO_3$ films[8-12] grown on various substrates.

The observed strain effect is usually interpreted qualitatively within the double-exchange model, since the hopping matrix element $t$ could be altered by epitaxial strain. Thus a compressive/tensile strain induces an increase/decrease in $T_{IM}$ by an increase/decrease in electron transfer due to the compressed/expanded Mn–O bond lengths[6,15]. However, recent detailed studies show that compressive strain does not always lead to enhancement of $T_C$[13], while the cationic vacancies due to the oxygen annealing significantly enhance the $T_C$ values much higher than any bulk values in the series compounds[5,14]. In most cases, tensile strain suppresses ferromagnetism and reduces $T_C$ in manganite films. But some anomalous results have also been reported, showing that $T_{C=}$is enhanced by tensile strain[18,24,25]. In a recent study Zhang et al (2001) [18] showed that a biaxial tensile strain also can lead to an enhancement of $T_{IM}$ in $La_{1-x}Ba_xMnO_3$ films due to orbital rearrangement. Nevertheless, the $T_{IM}$ enhancement mechanism based on orbital rearrangements has not been expected to cause an increase of $T_{IM}$ in the $La_{1-x}Ca_xMnO_3$ system[19]. It has been also proposed that the Jahn-Teller electron-phonon coupling plays an important role in the strain effect on $T_C$ [24]. Most interestingly, there are reports of multiple phase segregation in strained epitaxial films[17]. The ferromagnetic coupling within the metallic regions accounts for the changes of $T_C$ and conductivity.

These strain effects have been evaluated by the dependence of properties on the thickness and lattice matching between the films and substrates. Although consistent behaviors have been reported concerning the thickness dependence of $T_C$ of the manganite films[19,27,28] disagreement exists concerning the origin of the observed phenomena. Some investigators have argued that the difference in oxygen content is the most important factor responsible for the $T_C$ variation in manganites and that strain has less effect[12,28], while others have claimed that a change in structure, which is strongly coupled with the electronic system, must account for the origin of the behaviors observed [19,29]. Thus, the strain effect in manganite films has not become completely intelligible.

In the present work we have synthesized and studied the magnetotransport properties of $La_{0.7}Ca_{0.3}MnO_3$ epitaxial films deposited on $LaAlO_3$, $SrTiO_3$ and MgO single crystal substrates by pulsed laser deposition. It is known that lattice strain due to film-substrate mismatch affect the properties for thin films (< 100 nm) and it influences the magnetotransport properties ($T_{IM}$, $T_C$ and MR) significantly. However, for thicker films (> 100 nm) lattice strain gets relaxed and produces various kinds of imperfections such as stacking faults, dislocations etc. In strain relaxed epitaxial thin films of $La_{0.7}Ca_{0.3}MnO_3$ significant low field MR has been observed and this is supposed to be closely related to the degree of lattice relaxation and hence density of defects that disrupt the long range order [30,34]. In the present work the possible correlation between the degree of strain relaxation and the various physical properties such as $T_{IM}$, $T_C$ and low field MR has been investigated for relatively thicker films (~ 200 nm) of $La_{0.7}Ca_{0.3}MnO_3$ deposited by Pulsed Laser Deposition.

## 2. Experimental Details

### 2.1. Preparation of ceramic targets

For the preparation of the film by Pulsed Laser Deposition (PLD), first a stoichiometric ceramic target has to be synthesized. The $La_{0.7}Ca_{0.3}MnO_3$ target has been prepared by standard solid-state reaction of appropriate mixtures of 3N pure $La_2O_3$, CaO and $MnCO_3$. The powders are mixed thoroughly and calcined at 900°C for 12 hours in air. This process of mixing and calcinations has been repeated several times, each time the temperature is raised by 50 to 100°C up to 1100°C. After that the powder is pressed into a round pellet (target). The diameter of the targets is 20 mm, while the thickness is about 3 mm. The target is then sintered in air at 1300°C for 36 hours, and finally annealed in flowing oxygen at 1000°C for 12 hours. This produces very dense and hard targets so that particulate contamination of the laser-deposited films is reduced as much as possible.

## 2.2 Deposition of Films

The procedure we have used to grow thin manganite films is as follows. Prior to deposition, the substrates (for present study $LaAlO_3$, $SrTiO_3$ and MgO; all in (100) orientation) have been cleaned successively twice in acetone and isopropanol to remove dust or any other greasy contaminations. All three substrates were then glued side-by-side on the heater with silver paste to ensure good thermal contact and to have identical synthesis condition. The deposition chamber is evacuated down to background pressure of ~ $10^{-5}$ mbar or less. Then the vacuum pump is switched to a lower speed and the oxygen inlet, controlled manually by a needle valve, is opened to reach a constant, dynamic oxygen pressure (i.e. the deposition pressure) of 200-400 bars. A typical temperature for film deposition is 700-800 °C, the temperature being monitored by a thermocouple attached to the heater. In the meantime, the laser is switched on and brought to its maximum output power. After stabilization of pressure, temperature and laser power for about 15 minutes, the target is ablated during several minutes with the shutter closed in front of the heater. This is to remove all possible contaminations from the target surface without depositing material onto the substrate. The laser beam ($\lambda$ = 248 nm) is focused on a target with an energy density of ~2 J/cm$^2$ in a spot of 3-4 mm$^2$ with repetition rate (pulse frequency) of 2 Hz, the actual focus is set just in front of the target to minimize interaction between ejected ions and the laser beam. During the entire ablation process the deposition chamber is kept at a dynamic oxygen pressure of 200- 400 bar. The heater, carrying the substrate, is held at a distance of ~4 cm from the target. Films presented in this study are all deposited in on-axis geometry, i.e. perpendicular incidence of plume to substrate. The film thickness is monitored by the number of ablating laser pulses during deposition. The typical growth rate for the manganites at the parameters used is around 10 nm per minute. After the desired deposition time (typically around 20 minutes yielding a film thickness of around 200 nm), the laser is switched off, and the sample is kept in an oxygen atmosphere (usually 1 bar) at the deposition temperature (for • 1 h) and then slowly cooled (typically 5° /min) down to room temperature. After that films have been post annealed in flowing oxygen at 1000°C for 8 hrs to ensure full oxygen content.

### 2.3 Structural and Magnetotransport Characterizations

The crystal structure and orientation of all films were characterized by X-ray diffraction using $CuK_{\alpha}$ radiation ($\lambda$ =1.54106 Å) at room temperature in the $2\theta$ range of 20-80°. X-ray rocking curve analysis has been done to check the crystallinity of the films. The magnetization measurements have been performed in vibrating sample magnetometer at an applied magnetic field of 153 Oe in the range of 77-300 K. The electrical resistance has been measured in the range of 77-300 K by four-probe technique. The MR was calculated by using the resistance data with and without applied field ($H_{dc}$ = 1.0 T), which is applied parallel to film plane.

### 2.4 Results and Discussions

#### 2.4.1 Structure Analysis

XRD patterns of $La_{0.7}Ca_{0.3}MnO_3$ (LCMO) films fabricated on $SrTiO_3$ (001), $LaAlO_3$ have been found to match with the (00*l*) reflections expected from the bulk. The pattern confirms that the films are indeed single phase without any impurity and grown epitaxially on the c-axis as shown in Fig. 1. In order to study the effect of the substrate on the structure of the deposited films, the (002) peaks in the XRD pattern are analyzed.

The out of plane lattice parameters of the LCMO thin films deposited on STO, LAO and MgO single crystal substrates have been measured to be 3.878 Å, 3.889 Å and 3.899 Å respectively and the same for the bulk LCMO target is 3.92 Å. The in plane lattice parameters are found to be 3.890 Å, 3.911 Å and 3.912 Å respectively on STO, LAO and MgO substrates. Thus there is no appreciable difference between in plane and out of plane lattice parameters. The near equality of in plane and out of plane lattice parameters is due to the fact that the thickness of these films (~200 nm) is much larger than the critical thickness up to which the biaxial strain originating due to the difference between the lattice parameters of the substrate and LCMO can be effective. Since the film thickness is larger the lattice strain will get relaxed. At film thicknesses larger than the critical thickness, the lattice relaxation defined as $L = [(a_{film} - a_{substrate}) \times 100]/a_{substrate}$, will play the dominant role. In fact it has been shown that at larger thickness such as ~200 nm, the lattice strain relaxes leading to generation of variety of defects [30]. The epitaxial nature of the films has also been confirmed by the rocking curve analysis of the films and rocking curves corresponding to the (002) reflection is shown in Fig. 2. As seen in the figure, in the case of LCMO/STO film the intensity is highest and decreases for LCMO/LAO and LCMO/MgO films in that order. Similarly the FWHM is observed to be the minimum for the LCMO/STO film and increases for LCMO/LAO and LCMO/MgO in that order. The FWHM for these films is 0.39°, 0.52° and 1.24° respectively.

The above observed broadening may be due to one or more of the following factors (i) the mosaic spread, (ii) reduction of the long-range order in the lattice and (iii) increased static disorder due to imperfect growth. The long range order can be disrupted by the increased density of the stacking faults and tilting out of the in plane grains. Static disorders may lead to the effects such as electron localization etc. In the present case the magnitude of lattice relaxation has been found to be 0.384 3.057 and 6.411 percent for films deposited on STO, LAO and MgO single crystal substrates respectively[30]. Obviously as the difference between the lattice parameters of the LCMO film and substrates increases the magnitude of lattice relaxation also increases. At the same time the FWHM of the rocking curves is found to increase with the lattice relaxation. Increasing FWHM as evidenced by the broadened rocking curves with increasing lattice relaxation shows that defects are being generated as a consequence of this relaxation. As mentioned earlier the lattice relaxation indeed gives rise to extrinsic distortions/defects such as dislocations, grain-boundaries, stacking faults, cationic vacancies etc. The lattice relaxation, may not appreciably affect the more fundamental features such as the magnetic exchange interactions, but it may affect Jahn Teller distortions and hence the electron lattice coupling. Thus the effects of lattice relaxation will get reflected in the magneto-transport characteristics[7,9,18,19,22,23,28].

**2.4.2 Magnetotransport Properties**

The paramagnetic–ferromagnetic phase transition ($T_C$) was studied by measuring the magnetization ($M$) in the temperature range 300-77 K and the $M$-$T$ data is plotted in Fig. 3a. The PM – FM transition temperature or the Curie temperature ($T_C$) was determined from the maxima in the magnitude of the $dM/dT$. The observed $T_C$ values are ~ 244 K, 218 K and 175 K respectively for LCMO films deposited on STO, LAO and MgO. The $T_C$ of the bulk target has been measured to be ~ 245 K and the corresponding $M$ – T data is plotted in Fig. 3b. The $T_C$ of the LCMO bulk target and the LCMO thin film on STO substrate are nearly equal while that of the LCMO films on LAO and MgO substrates are smaller than that of the bulk LCMO target. In fact, the $M$ – $T$ data shows that the transitions in the LCMO thin films on LAO and MgO substrates are not as pronounced as in case of the LCMO bulk and thin film on STO. The broadness of the transition suggests towards either a percolative regime or a fluctuation regime having its origin in competition between related interactions around $T_C$. There are host of factors that can account for a $T_C$ depression as observed in the LCMO thin films grown on LAO and MgO substrates. One of such possible factors is the biaxial lattice strain, but as the film thickness is around ~ 200 nm, the lattice strain is expected to be almost completely relaxed. Therefore the contribution from this factor will be negligibly small[7,9,18,19,22,23,28]. The second factor leading to $T_C$ depression is the oxygen deficiency and it may well be one of the reasons leading to $T_C$ depression in the present case, especially in the case of films deposited on MgO substrates. Another reason and perhaps the dominant one in the present study might well be the structural disorders originating due the lattice strain relaxation. In fact as can be seen in Table A.1, in case of LCMO film on STO where the magnitude of lattice relaxation is very small (0.384 %), the $T_C$ of the film is nearly same as that of the bulk.

However, as the magnitude of the lattice relaxation increases to 3.057 and 6.411 percent for LCMO films on LAO and MgO, the $T_C$ decreases quite rapidly to 218 K and 186 K, respectively. The $T_C$ depression can be accounted as in the following: As pointed out and discussed earlier, due to the difference between the lattice parameters of the substrate and film, lattice relaxation occurs. This may lead to generation of various types of defects, such as mosaic spread and decrease in the long range order due to increased density of stacking faults, in plane tilting of grains, creation of oxygen vacancies etc. It is expected that the density of such disorders would increase as the magnitude of the lattice relaxation increases. Induced disorders that disrupt the long range order in the lattice, such as the stacking faults may produce modifications in the Jahn Teller distortion of the $MnO_6$ octahedra leading to electron localization effects and hence may increase the polaronic component to the transport in the paramagnetic phase. The larger value of the lattice relaxation in LCMO/MgO film thus results in the higher density of defects such as the stacking faults, dislocations, etc. The increased density of these lattice defects may lead to larger distortion of the local lattice environment such as the $MnO_6$ octahedra.

The temperature dependence of resistivity of all the LCMO thin films was measured between 300K and 77K and the corresponding data are plotted in Fig. 4. At room temperature (~300K) the resistivity of LCMO films on STO, LAO and MgO are measured to be 36 mΩ-cm, 31 mΩ-cm and 38 mΩ-cm. Thus at room temperature all the films, irrespective of the substrate have nearly the same resistivity value. However, as the temperature is reduced resistivities are observed to increase and at temperature well below room temperature insulator-metal transition ($T_{IM}$) is observed. The $T_{IM}$ s of LCMO films on STO, LAO and MgO are found to be 243K, 217K and 191K and the corresponding resistivities at $T_{IM}$ are 72 mΩ-cm, 113 mΩ-cm and 275 mΩ-cm, respectively. Thus compared to the room temperature values the resistivity of LCMO films on STO, LAO and MgO undergo a 2-fold, 3.65-fold and 7.25-fold while the $T_{IM}$ for the LCMO film on MgO is higher than the $T_C$. At 77 K the resistivity of all the films have been measured to be less than 2 mΩ-cm. Thus the observed variation of the resistivity of the LCMO films shows that increased lattice relaxation enhances the resistivity at $T_{IM}$ while the room temperature as well as the low temperature resistivity remain nearly independent of the nature of the substrate and hence the lattice relaxation. The resistivity enhancement around $T_{IM}$ and in the paramagnetic phase in case of LCMO film having larger lattice relaxation suggests increased in carrier localization.

### 2.4.3 Nature of Transport

As outlined in the above, disorders are invariably present in the present films, the ρ-$T$ data above $T_C$ was analyzed in the framework of the Mott's variable range hopping (VRH) of polarons which is given by [31],

$$\rho(T) = \rho_0 \, exp \left(\frac{T_0}{T}\right)^{1/4}$$

Where $\rho_0$ is the residual resistivity and $T_0$ is the characteristic VRH temperature. It has been observed that in all LCMO films the data above $T_C$ fits well to the above Mott's VRH equation. This is amply clear in the ln (ρ) - $T^{0.25}$ plots for the LCMO films as depicted in Fig. 5 (a,b,c).

The value of the constant ($T_0$) at room temperature has been evaluated from the linear fit of the resistivity data and for LCMO films an STO, LAO and MgO the values of $T_0$ are found to be 1.24 x $10^7$ K, 2.25 x $10^7$ K and 3.83 x $10^7$ K respectively. Thus with increased lattice relaxation the characteristic VRH temperature ($T_0$) also increases. Since the value of the parameter $T_0$ is a measure of the strength of the Jahn-Teller distortion and is inversely related to the extent of the localized states, therefore, the increasing value of $T_0$ suggests that increased lattice relaxation decreases the localization length which in turn reduces the average hopping distance.

The localization length (1/ α) has been calculated using the modified formula proposed by Viret et al. (1997) [32]. They have proposed that in case of manganites with 30% divalent cation doping at the rare earth site, the carrier localization above $T_C$ is caused by a random potential of magnetic origin. This potential is due to the Hund's rule coupling -$J_H$ $\vec{S}_i . \vec{S}_j$ between localized Mn $t_{2g}$ ion cores (S=3/2) and spin s of $e_g$ electrons in the conduction band. The modified formula for localization length ($1/\alpha$) in this model is given by,

$$\frac{1}{\alpha} = \left( \frac{171\ U_m V}{K_B T_0} \right)^{1/3}$$

Here $U_m = 3\frac{J_H}{2}$ is the splitting between the spin up and spin down $e_g$ bands and its value has been found from optical spectra to be 2eV [33]. V is the unit cell volume per Mn ion and $T_0$ is the characteristic VRH temperature and $k_B$ is the Boltzmann constant. Substituting the value of $U_m$ and $k_B$ above expression becomes,

$$\frac{1}{\alpha} = \left( \frac{3.965\ x 10^{6} V}{T_o} \right)^{1/3}$$

Thus the average nearest-neighbor hopping (R) distance is given by

$$R = \left( \frac{9}{8\pi\alpha\ N(E) K_B T} \right)^{1/4}$$

where N (E) is the density of available states in the random potential regime of Viret et al. [33] and its value estimated by them for LCMO is 9 x $10^{26}$ $m^{-3}eV^{-1}$. Using this value of N (E) and the Boltzmann constant the above expression transform to,

$$R = \left( \frac{4.6112\ x\ 10^{-24}}{\alpha T} \right)^{1/4}$$

Using the above expressions the value of the localization length $1/\alpha$ has been calculated to be 2.65Å, 2.18Å and 1.83Å for LCMO films on STO, LAO and MgO respectively. Similarly, the average hopping distance R has been found to be 14.21Å, 13.53Å and 12.98Å for LCMO films grown on STO, LAO and MgO, respectively. From these data of localization length ($1/\alpha$) and average hopping distance (R), it becomes evident that as the lattice relaxation increases in the sequence STO (-0.384)→LAO (3.057) →MgO (-6.411) the localization length sequentially decreases, that is, the carriers becomes more and more localized resulting a decreases in the average hopping distance. This also suggests that the lattice relaxation indeed leads to generation of defects such as stacking faults, dislocations and other defects. The decrease in the localization length (1/α) and average hopping distance (R) is quite significant at 4.8 % and 8.7 % as compared to the same for the LCMO thin film on STO.

The low field magnetoresistance (LFMR) of the LCMO thin films grown of different substrates was measured in the temperature range 300 – 77 K and at magnetic fields $H_{dc} \leq 1.0$ T and the corresponding data are plotted in Fig. 6a and Fig. 6b. It is observed that at $H_{dc} = 0.5$ T as well as $H_{dc} = 1.0$ T the MR becomes significant around $T_C$ while the peak in the MR occurs at a lower temperature. This feature, as also pointed out by de Andres et al. [34], is in sharp contrast to the high field MR that peaks around $T_C$. For the LCMO thin film grown on STO the low field MR becomes significant at $T_C$ while the peak MR of ~7 % and ~18 % respectively at $H_{dc}$ = 0.5 and 1.0 T occurs at a lower temperature around T ~ 220 K. The same behaviour is observed in case of films on LAO and MgO although the MR values in this case are more than double the corresponding values measured for LCMO film on STO. However, apart from the magnitude, the temperature spread of MR around the respective peak MR values also increases as one moves from film on STO (low lattice relaxation) to film on MgO (high lattice relaxation). The LCMO film on STO by virtue of low lattice relaxation (~ 0.4 %) is expected to have relatively smaller density of extrinsic defects while the films on LAO (Lattice relaxation ~ 3.06 %) and MgO (Lattice relaxation ~ 6.411 %) have much higher density of disorders and defects. It has been discussed by de Andres et al. [34] that in epitaxial thin films the low field MR around $T_C$ or $T_{IM}$ is related to the lattice distortions that could result as a consequence of the lattice relaxation induced defects such as the stacking faults etc. In the present investigation it has been shown from the rocking curve analysis as well as the temperature dependence of resistivity in the paramagnetic state that lattice distortions indeed increase with increasing lattice relaxation. It is known that around $T_C$ polaron clusters form in the FM phase of the thirty percent doped LCMO. This polaron cluster formation is due to a drastic reduction in the bandwidth due to the lattice distortions around the defects. Since in the present investigation the density of defects increases as the lattice relaxation increases, therefore the polaraonic density is also expected to increase as one move from STO to LAO and then to MgO. These polarons are weakly bound in the FM phase and consequently can be delocalized by a relatively weak magnetic field. Thus the low field MR in epitaxial thin films around $T_C$ or $T_{IM}$ has its origin in the delocalization of the small polaron clusters formed around the $T_C$ or $T_{IM}$ in the FM phase.

The contribution of the defects/disorders is more or less confined to the temperature regime around $T_C$ and $T_{IM}$. In the LCMO films, polaron clusters form in the ferromagnetic phase around $T_C$ by a drastic reduction of the bandwidth due to the lattice distortions around the defects. The decrease in the bandwidth caused by lattice distortion is given the Narimanov-Varma [35] formula

$$W = W_0 \cos\frac{\theta}{2} \exp - \frac{k(\delta u)^2}{\hbar\omega_0}$$

Here θ is the angle between the two Mn spins and $\delta u$ are lattice distortions that can be dynamic (e. g. phonons), intrinsic (related to La substitution by a divalent cation) or due to lattice defects around impurities. The lattice distortions around the defects favour the localization of polarons ($T_0$ increases). As proposed by de Andres et al. [34] these polarons are rather weakly bond in the ferromagnetic phase and therefore can be delocalized by a relatively smaller dc magnetic field. These small polarons can only be formed near $T_C$ when the carrier bandwidth is reduced either by spin disorder or by lattice fluctuations. In the present case this is quite obvious from the decrease in the $T_{IM}$ values with increasing lattice relaxation and also from the fact that the value of the parameter $T_0$ that eventually determines that localization length increases as the magnitude of the lattice relaxation increases in the LCMO films on STO, LAO and MgO.

**3. Conclusions**

We have synthesized epitaxial $La_{0.7}Ca_{0.3}MnO_3$ films on $LaAlO_3$, $SrTiO_3$ and MgO substrates and investigated the effect of strain relaxation on the magnetotransport properties of these films. It has been found that, all the films are epitaxial, single phase and have cubic unit cell. The out of plane lattice parameter are 3.878Å, 3.889 Å and 3.899Å respectively for films on STO, LAO and MgO. The FWHM as evaluated through rocking curve (ω scan) comes out to be 0.39° , 0.52° and 1.24° for films on STO, LAO and MgO respectively. As the films investigated in the present case are thicker (~200 nm), the substrate induced strain does not persist but it relaxes and these relaxation gives rise to extrinsic distortions/defects such as dislocations, grain-boundaries, stacking faults, cationic vacancies etc. The lattice relaxation has been found to be 0.384 3.057 and 6.411 percent for films deposited on STO, LAO and MgO. The observed $T_C$ values are ~ 244 K, 218 K and 186 K respectively for LCMO films deposited on STO, LAO and MgO. The $T_{IM}$ s of LCMO films on STO, LAO and MgO are found to be ~ 243K, 217K and 191K. The decrease in $T_{IM}$ and $T_C$ for MgO films have been explained on the basis lattice strain relaxation.

Higher degree of relaxation creates more defects which affect the magnetotransport properties and $T_{IM}$ ($T_C$) decreases. Keeping in view the variable presence of disorder in the present films, we have analyzed the transport above $T_C$ through Mott`s VRH model. Based on this model the increase in lattice relaxation will produce defects, which will result in decrease of the tendency of charge localization (hopping distance also decreases) and so will $T_{IM}$ and $T_C$ decreases. As already described in section 2.4.3 these are in keeping with our experimental results.

**Acknowledgements:** This work was supported financially by UGC and CSIR, New Delhi. One of the author (PKS) acknowledges CSIR, New Delhi for the award of SRF. Authors are also thankful to Professors A R Verma, C N R Rao, T V Ramakrishnan, Vikram Kumar, S B Ogale, A K Raychaudhari ,Dr Kishan Lal and Dr. N. Khare for valuable discussions. We also thanks Dr. Alok Banarjee for helping in magnetization measurements.

| **Characteristics** | **STO** ( 3.905 Å) | **LAO** ( 3.821 Å) | **MgO** ( 4.216 Å) |
|---|---|---|---|
| Out of plane lattice constant | 3.878 Å | 3.889 Å | 3.899 Å |
| Lattice Relaxation $L = (a_F - a_S) x100/a_S$ | - 0.384 | 3.057 | -6.411 |
| $T_C$ | 245 K | 220 K | 186 K |
| $T_{IM}$ | 243 K | 217 K | 191 K |
| ρ at 300K and at $T_{IM}$ | 36 mΩ-cm, 72 mΩ-cm | 31 mΩ-cm 113 mΩ-cm | 38 mΩ-cm 275 mΩ-cm |
| $\rho(T_{IM})/\rho(300\ K)$ | 2 | 3.65 | 7.25 |
| M R % (0.5T) (At $T \sim T_C$, $T_{IM}$ and 80K) | 2.5, 3.5, 1.4 | 7.1, 11.2, 0.3 | 13.9, 21, 3.5 |
| M R % (1.0T) (At $T \sim T_C$, $T_{IM}$ and 80K) | 9.1, 12.5, 2.1 | 16.5, 5.9,4.4 | 25.7, 38, 4.7 |
| $MR_{max.}$ (0.5 T, 1.0 T) | 6.9, 17.6 | 22.5, 41 | 26, 44 |
| T ($MR_{max.}$) (0.5 T, 1.0 T) | 220, 225 K | 195, 200 K | 175, 180 K |
| $T_0$ | $1.242 \times 10^7$ K | $2.251 \times 10^7$ K | $3.830 \times 10^7$ K |

**Table 1:** The various characteristics parameters of epitaxial films of $La_{0.7}Ca_{0.3}MnO_3$ $SrTiO_3$ (001), $LaAlO_3$ (001) and MgO (001)

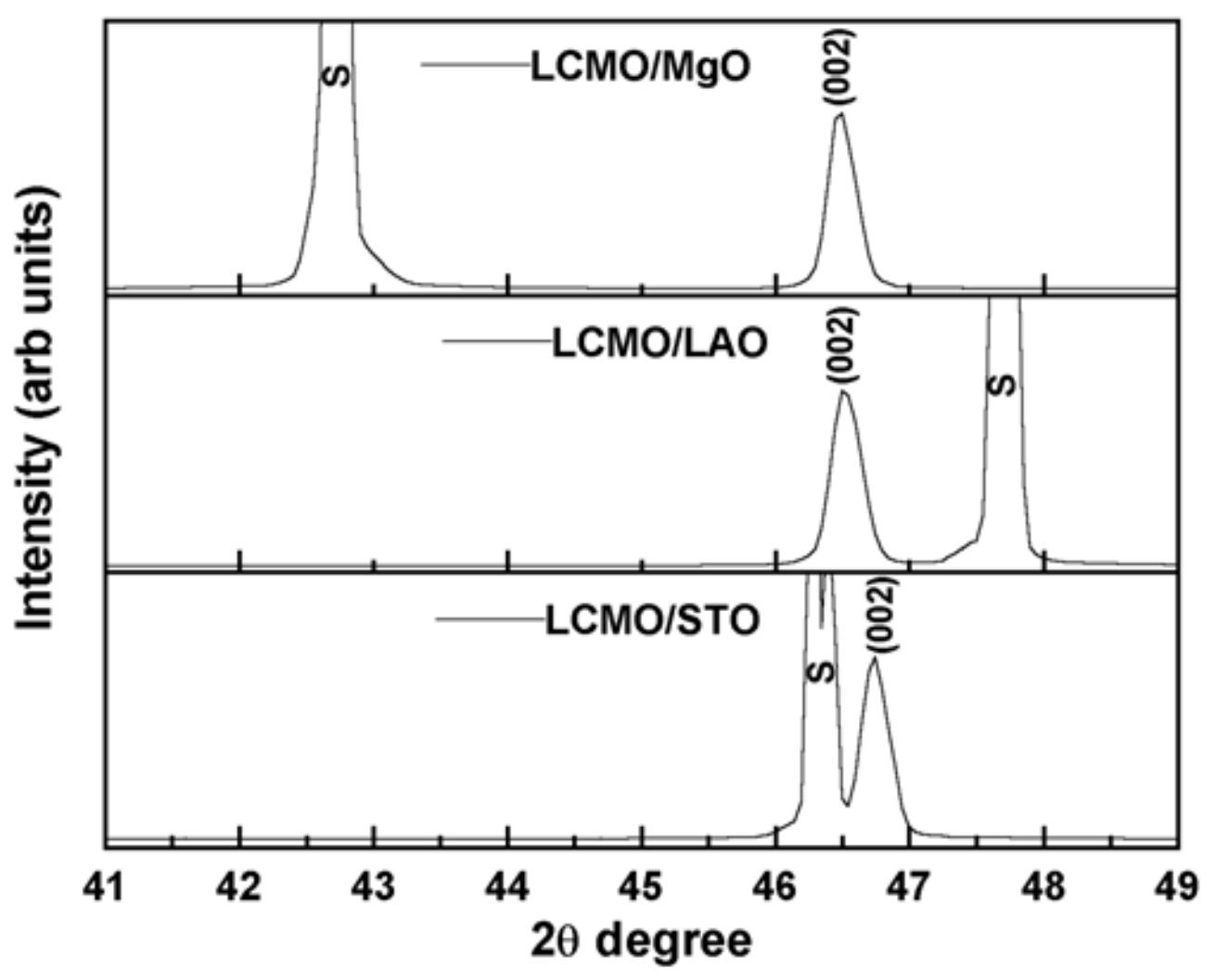

**Figure 1:** XRD pattern of pulsed laser deposited epitaxial films of $La_{0.7}Ca_{0.3}MnO_3$ (LCMO) on STO, LAO and MgO substrates.

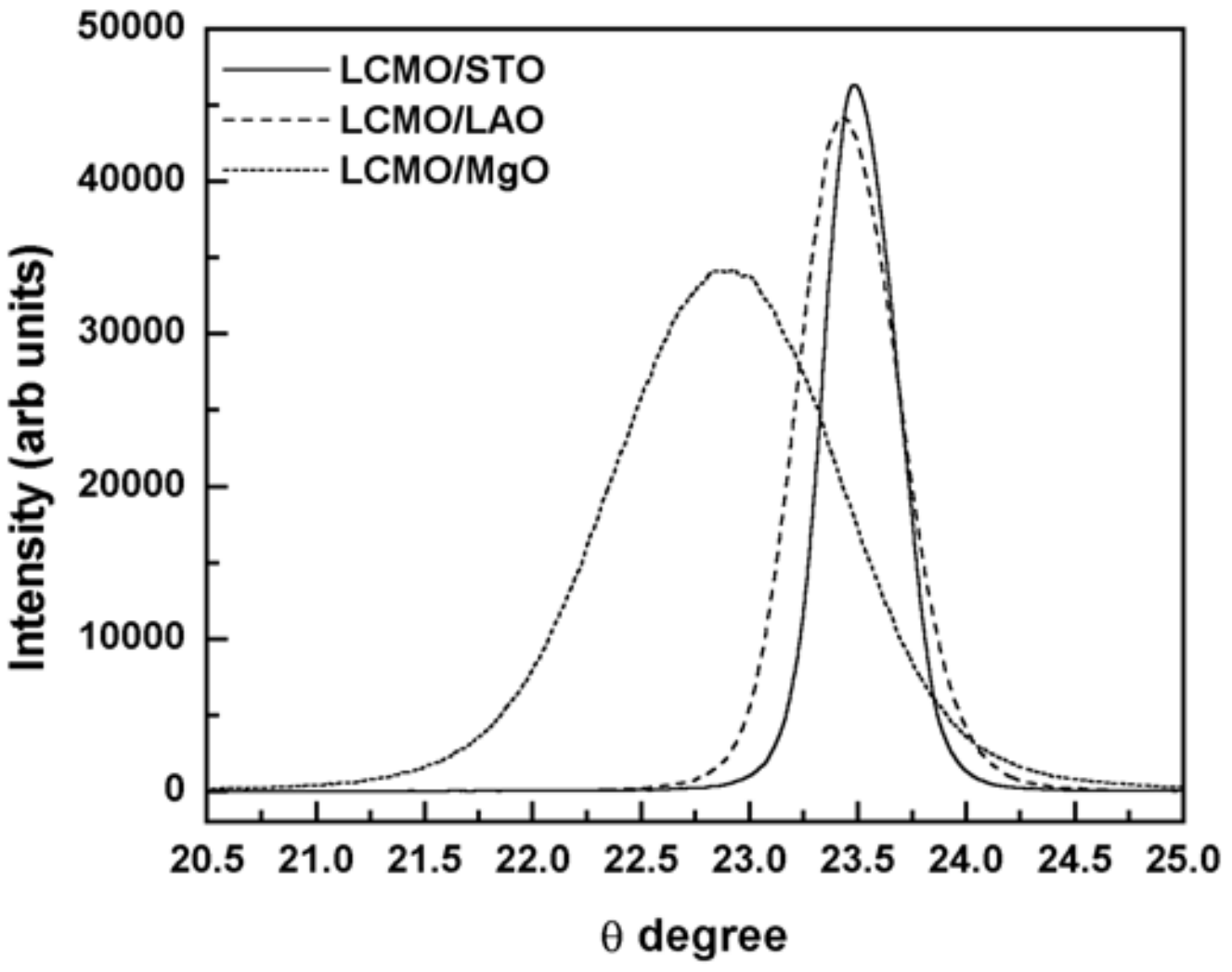

**Figure 2:** ω scan of the (002) reflection for LCMO films on STO, LAO and MgO.

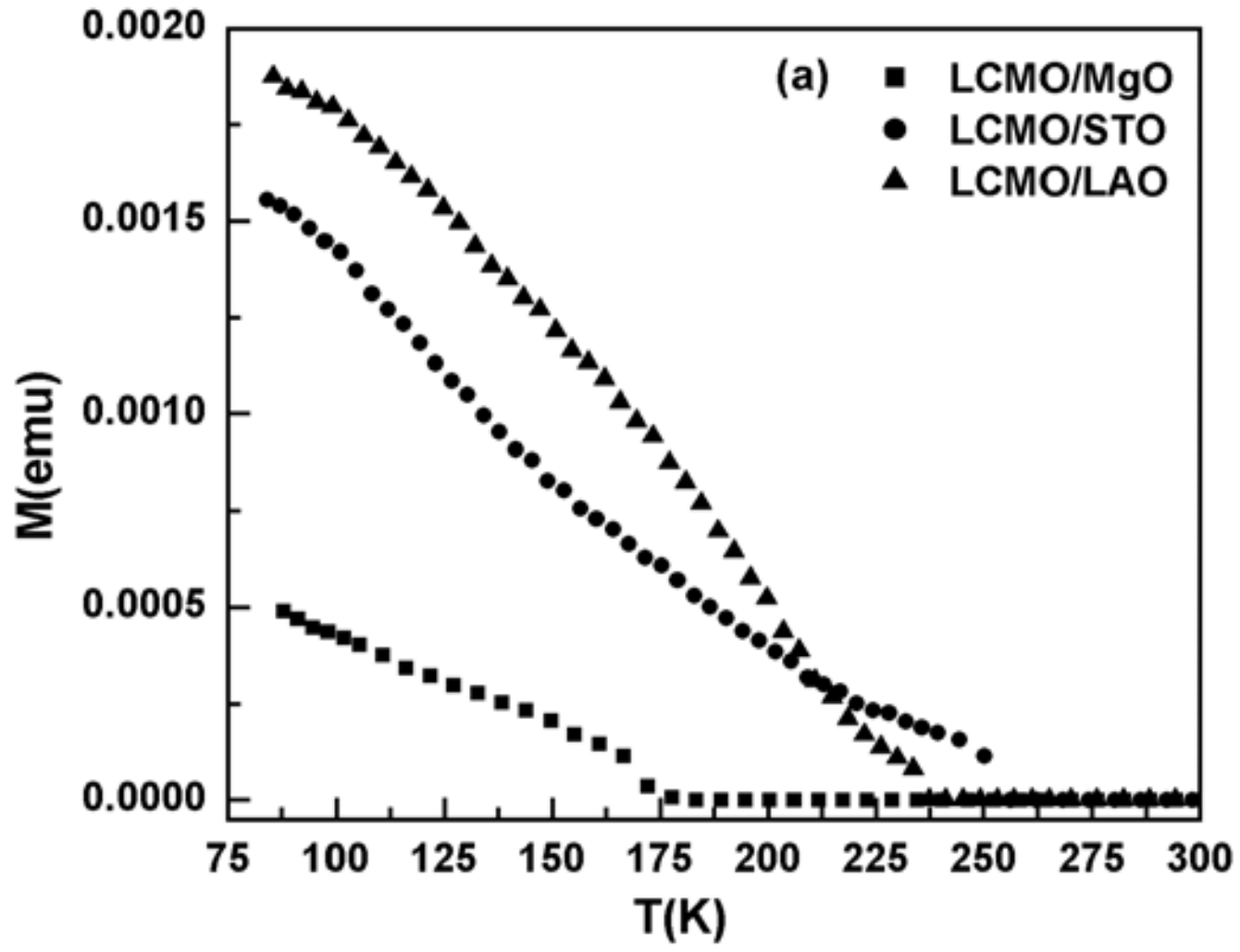

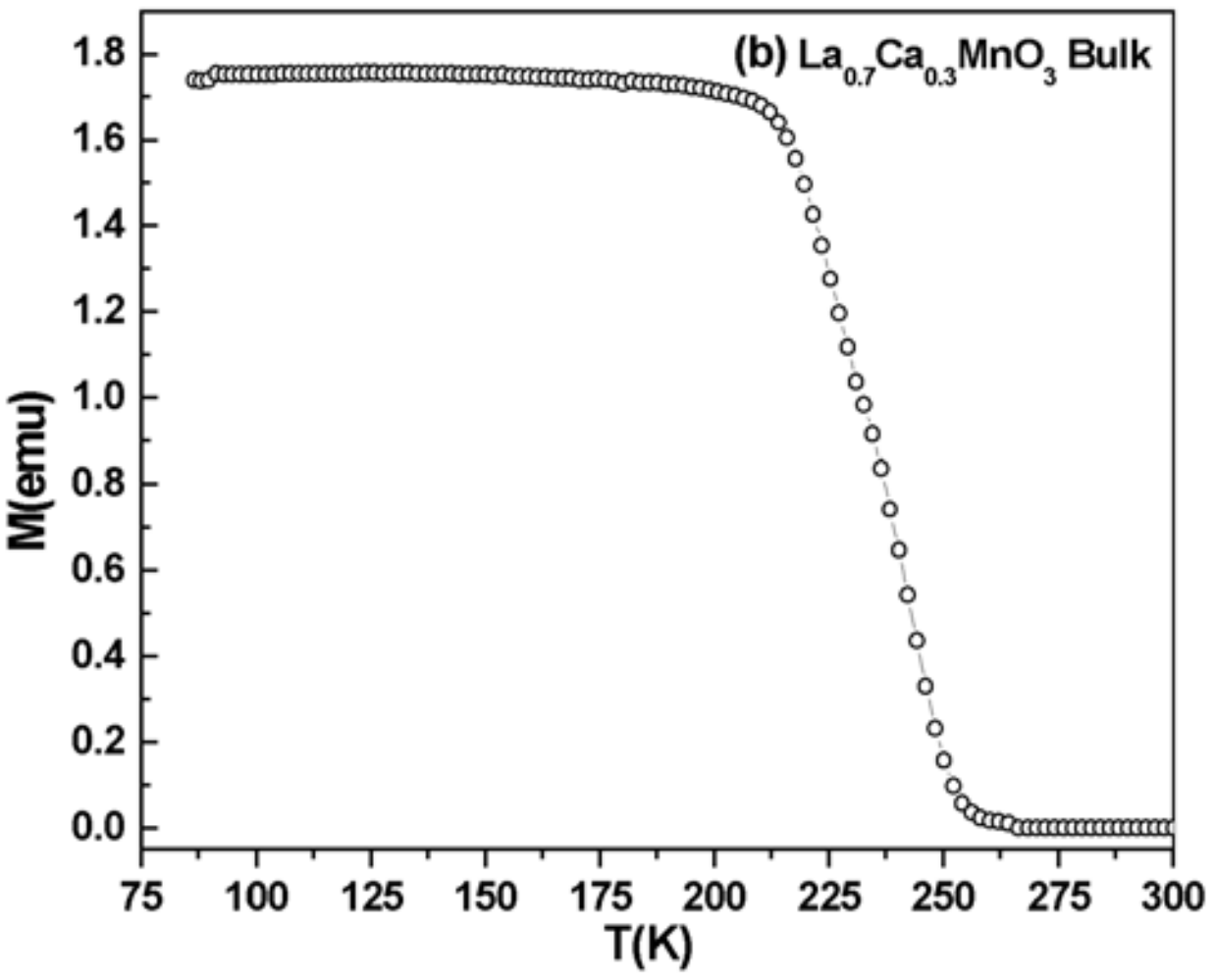

**Figure 3:** Temperature dependence of the magnetization for (a) LCMO films on STO, LAO and MgO substrates and (b) bulk $La_{0.7}Ca_{0.3}MnO_3$.

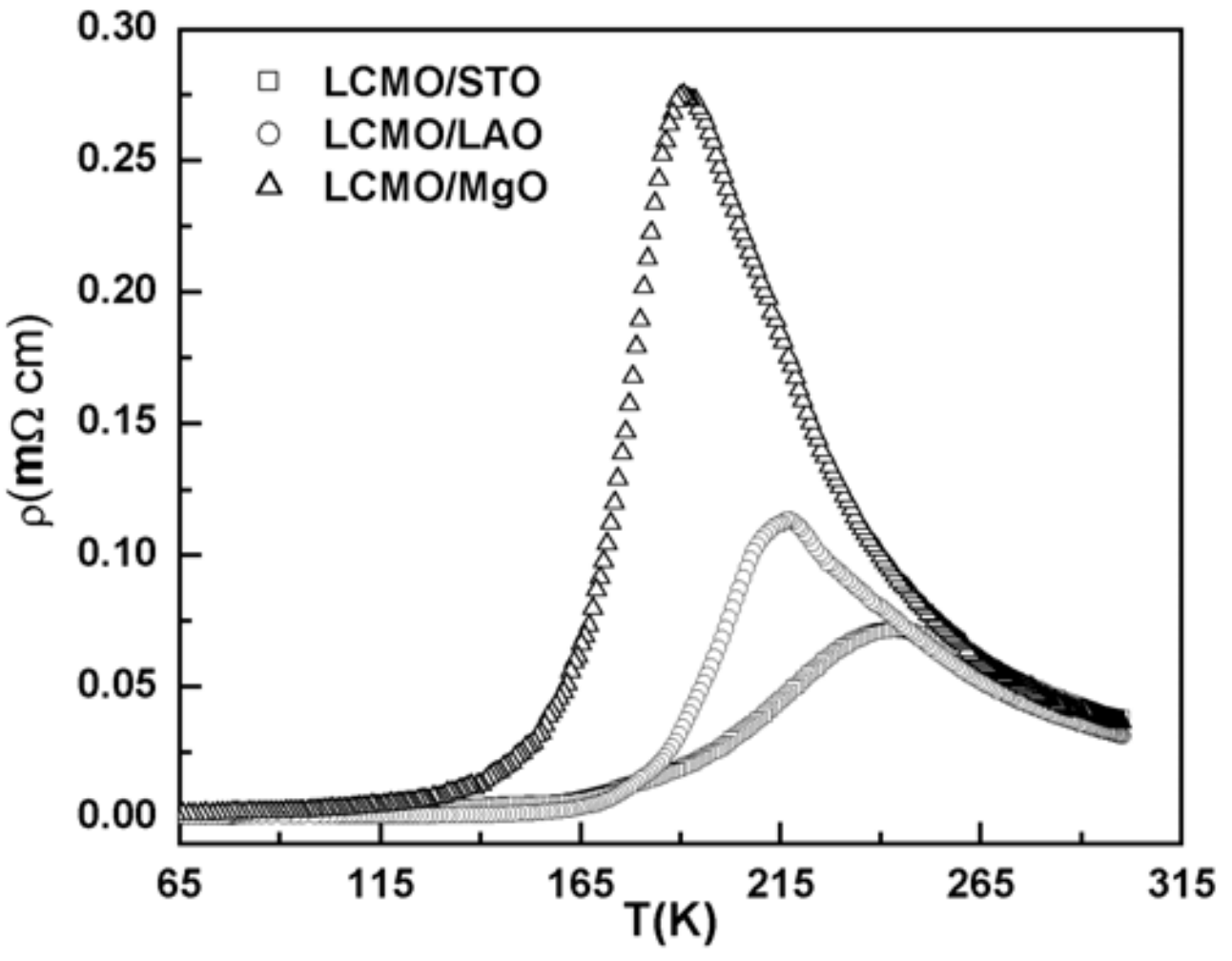

**Figure 4:** Typical temperature dependence of electrical resistivity for LCMO films on STO, LAO and MgO.

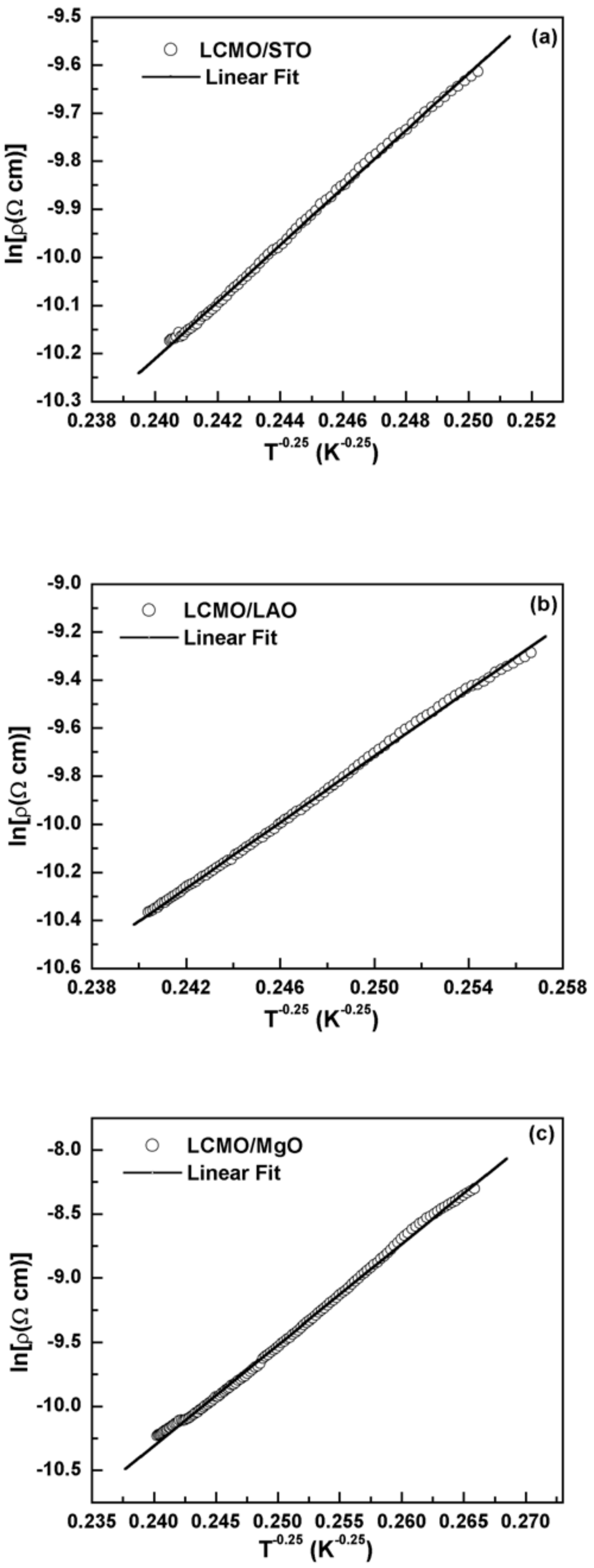

Figure 5: Variation of ln (ρ) with $T^{0.25}$ for LCMO thin films on $SrTiO_3$ , $LaAlO_3$ and MgO substrates.The open circle shows the experimental data and the solid line is the best linear fit.

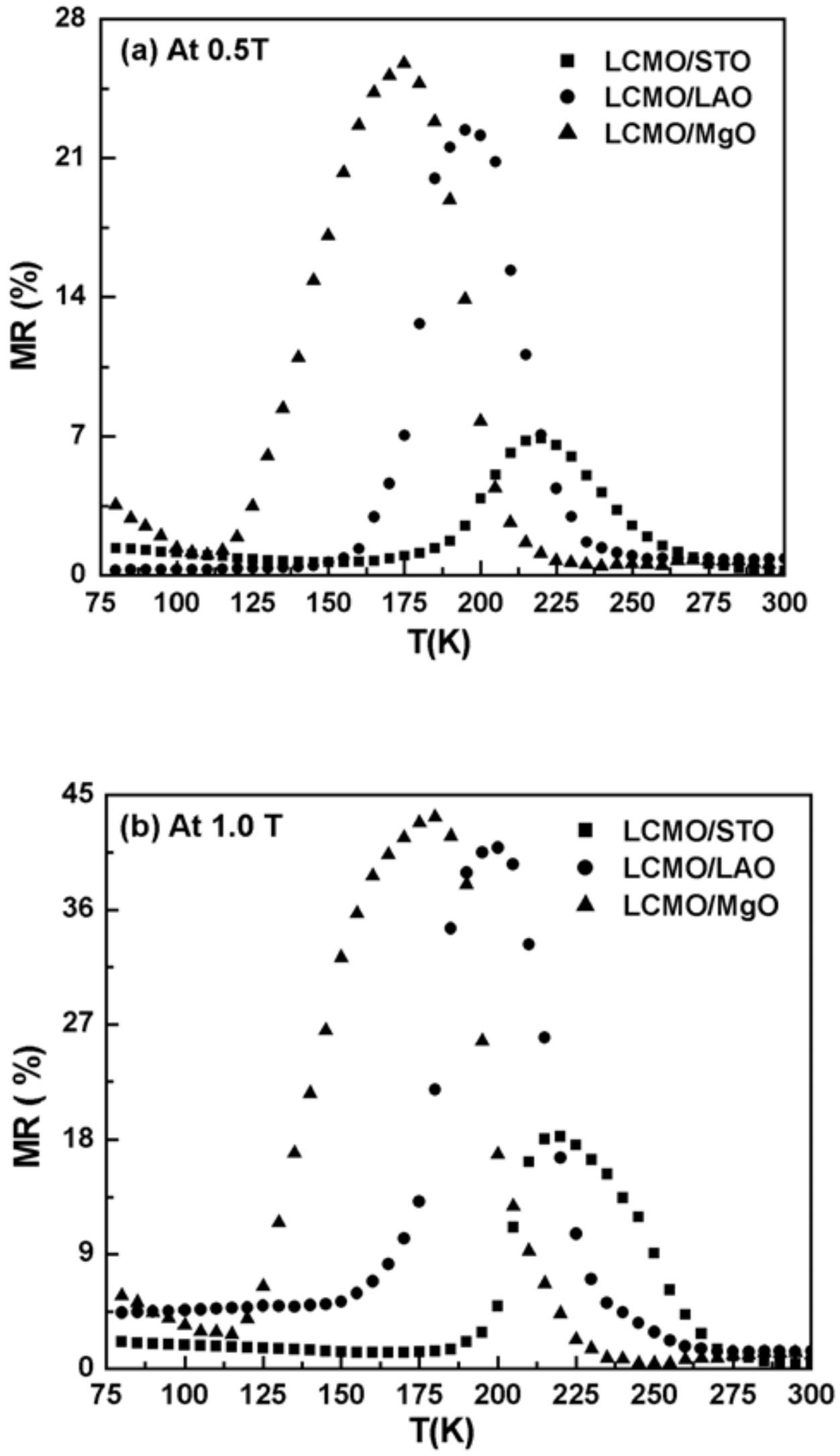

**Figure 6:** The temperature dependence of MR for LCMO films in an applied magnetic field of (a) 0.5 T and (b) 1.0 T.